# Possible superlattice formation in high-temperature treated carbonaceous MgB$_2$ at elevated pressure

Oliver Tschauner[1], Daniel Errandonea[2,3], and George Serghiou[4]

1: High Pressure Science and Engineering Center, Department of Physics, University of Nevada, 4505 Maryland Parkway, Las Vegas, Nevada 89154-4002, USA.
2: Departamento de Física Aplicada-ICMUV, Universidad de Valencia, Moliner 50, Burjassot, E 46100, Valencia, Spain.
3: HPCAT, Advanced Photon Source, Argonne National Laboratory, 9700 S. Cass Avenue, Chicago, Illinois 60439, USA.
4: School of Engineering and Electronics, University of Edinburgh, Mayfield Road, EH9 3JL UK.

We report indications of a phase transition in carbonaceous MgB$_2$ above 9 GPa at 300 K after stress relaxation by laser heating. The transition was detected using Raman spectroscopy and X-ray diffraction. The observed changes are consistent with a second-order structural transition involving a doubling of the unit cell along *c* and a reduction of the boron site symmetry. Moreover, the Raman spectra suggest a reduction in electron-phonon coupling in the slightly modified MgB$_2$ structure consistent with the previously proposed topological transition in MgB$_2$. However, further attributes including deviatoric stress, lattice defects, and compositional variation may play an important role in the observed phenomena.

**Introduction**

The recently reported superconductivity in MgB$_2$ [1] has been described by the BCS-mechanism [2] in a system with two almost independent bands of pair forming electrons [3 - 8]. *Ab initio* calculations indicate that the size of the larger of the two superconducting gaps depends markedly on crystallographic direction [9]. The phonon of E$_{2g}$ symmetry involved in the Fröhlich interaction with the two-dimensionally confined boron σ-electron bands has been found to be extremely anharmonic, and the coupling itself to be nonlinear [10]. This coupling induces the larger of the two superconducting gaps and is responsible for the superconducting transition around 40 K. Hence, static lattice distortions in the boron layer are expected to strongly affect superconductivity in MgB$_2$. Structural modifications of the basic MgB$_2$ structure then can open synthetic paths for new structurally related superconductors with high T$_c$. However, so far any reported modification was induced by chemical changes and has led to a reduction in T$_c$.

Here we report the first direct experimental findings supporting such a modification for carbonaceous MgB$_2$ at high pressure. By combining laser heating in diamond-anvil cells (DAC's), Raman spectroscopy, and X-ray diffraction, we present indications for a transition from space group P6/mmm to a subgroup of P6$_3$mc at around 9 to 10 GPa and 300 K after stress relaxation by laser heating to above 1500 K. The P6/mmm structure of MgB$_2$ is composed of planar hexagonal arrays of B atoms intercalated by Mg atoms such that each Mg atom has twelve B atoms in equal distance as nearest neighbours, while each B is surrounded by six equidistant Mg atoms. Hence, the Mg atoms reside at a distance of ½ *c* above the centres of the regular six-fold arrays of B atoms. All Mg atoms reside on 1a-sites and all B atoms assume the Wyckoff position 2d. On the other hand, in the proposed new phase of MgB$_2$ the MgB$_2$ cell is doubled and the B atoms exhibit reduced site-symmetry.

**Experimental**

Micron-sized MgB$_2$ powder (analyzed for carbon (C) and iron (Fe), with 0.38 at % C, and 0.16 at % Fe, obtained from Alfa-Aesar) was characterised by Raman spectroscopy and X-ray diffraction at ambient conditions. There was no indication of additional phases in the starting material. The cell parameters were *a* = 3.086(5) Å and *c* = 3.521(4) Å, which is in excellent agreement with previous studies [11,12].

The Raman spectra of the starting material show a rather hard $E_{2g}$ phonon around 670 cm$^{-1}$ and a strong density-of-state peak centred around 694 cm$^{-1}$. The high energy of the $E_{2g}$ phonon indicates reduced electron-phonon coupling. Hardening of this phonon and reduction of its coupling to the B σ-electron bands is expected for carbonaceous $MgB_2$ and has been reported [13, 14], although the absolute energies of both peaks in our sample material are higher than reported in [13]. This difference and the enhanced density-of-state scattering around 694 cm$^{-1}$ point to additional loss of phonon coherence e.g. from an enhanced defect density as by deviations from stoichiometry [15,16,17].

We loaded pre-pressed pellets of the sample powder in (DAC's) using NaCl and argon (Ar) as pressure transmitting media as well as no pressure medium in order to establish different regimes of deviatoric stress and provide a check for potential chemical interactions. The gasket material was rhenium (Re), and the sample chambers had diameters of 100 µm and heights of 30 to 40 µm. The samples embedded in NaCl and Ar were 50×50 µm$^2$ in size and 5 µm thick pellets of pre-pressed $MgB_2$ powder. All DAC's were pressurized to 9.7 GPa at room temperature (RT) as measured by ruby fluorimetry. At this pressure the samples were examined at 300 K by Raman spectroscopy both before and after laser heating. We used the micro-Raman spectrometer of GSECARS at Sector 13 of the APS-ANL synchrotron for collecting Raman spectra of the pressurized samples right before and after laser heating. We used the 514.5 nm excitation line of an Ar-ion laser operated at 200 mW power focussed down to 50 µm, a Spex 0.5 m monochromator with a 1200 grooves/mm grating, and a liquid nitrogen cooled CCD camera detector from Oxford Instruments. The starting material and the retrieved samples were characterized with an in-house micro-Raman spectrometer using 514.5 nm excitation wavelength of an Ar-ion laser run at 70 mW power with a 50% neutral density filter and the beam defocused to 500 µm in diameter. We used a Jobin-Yvon U1000 double axis monochromator and a liquid nitrogen cooled CCD camera detector from Oxford Instruments. We did not observe effects of local heating from the Ar-ion laser beam at 9 GPa, probably because the effective laser power density in the sample is lowered by attenuation and focal spot size distortion through the diamond anvil. Angle-dispersive X-ray diffraction patterns were collected in two independent experiments at the 16ID-B undulator beamline at the High Pressure Collaborative Access Team (HPCAT), sector 16 of the APS-ANL synchrotron, using monochromatic beams of 0.3681 Å or 0.3738 Å wavelength and MarCCD or Mar345 image plate detectors placed 280 mm and 326 mm away from the samples, respectively. The size of the monochromatic micro-focused X-ray beam was 12 μm × 14 μm achieved by using two Kirkpatrick-Baez mirrors and a 30 μm diameter Mo cleanup pinhole to eliminate the beam-tails. The diffraction images were integrated and corrected for distortions [18].

The samples were double-sided laser heated with the radiation of two Nd:YLF lasers (Photonics GS40, 85 W, $TEM_{01}$ mode, $\lambda = 1053$ nm) available at HPCAT [19]. The temperature was raised to just above the limit of visual observation and was measured spectroradiometrically to be between 1500 and 1800 K with a radial temperature gradient of less than 3% within the central 30 μm region. A detailed description of the laser-heating system has been published elsewhere [19]. All samples were heated for 30s. The beam on one side was scanned over the entire sample surface. The main goals of this heating procedure at high pressure are (a) to search for possible quenchable high temperature-high pressure phases of $MgB_2$, (b) to relax deviatoric stresses in the elastically anisotropic sample powder aggregate. After heating, the pressures were 9.5, 9.7, and 10.5 GPa for the samples heated in NaCl, Ar and without pressure medium, respectively. We collected diffraction data and Raman spectra after temperature quenching.

**Discussion and Results**

The Raman spectra of $MgB_2$ [3] and its analogues $AlB_2$ [3,17] and LiBC [20,21,22] provided several unexpected features: In $MgB_2$ the B-B in-plane stretching mode of $E_{2g}$ symmetry is substantially softened in comparison to $AlB_2$ and LiBC due to the strong electron-phonon coupling of this vibration in $MgB_2$. Upon compression and decompression of $MgB_2$ under non-hydrostatic conditions additional broad features occur in the Raman spectrum [15,17]. Consistent with this finding the Raman-spectra collected in our study before heating (fig 1a) show only a broad feature around 690 and a shoulder around 770 $cm^{-1}$. Under conditions closer to hydrostatic [23] only one peak occurs at 725 $cm^{-1}$ at this pressure. Spectra collected after heating (fig. 1b,c,d) are distinctly different: The characteristic Raman peak observed in unheated $MgB_2$ (fig. 1a, [17, 23]) disappeared and two peaks appear instead. The spectrum of $MgB_2$ heated without pressure medium shows two distinct peaks at 636 and 757 cm$^{-1}$ (fig. 1b), the spectrum of MgB$_2$ heated in NaCl shows two peaks also, but with a larger splitting (fig. 1c) [24]. These Raman spectra were collected on the same samples as the diffraction data (see below). In addition, we collected Raman spectra on MgB$_2$ heated in an Ar pressure medium (fig. 1d). In this spectrum there are again two peaks but with even greater separation in energy than in fig. 1c. In all three cases there are only two peaks in the Raman spectrum but there are significant differences in the energies of these two peaks. Further, there is a correlation of the energy difference between the new peaks and the amount of deviatoric stress, induced in the sample by rather low pressure gradients in annealed Argon to very high values in MgB$_2$ without pressure-medium, while NaCl provides an intermediate level of pressure gradients, being less soft than the van der Waals solid Argon but softer than MgB$_2$ which has a high bulk modulus [25, 26]. Hence, we suggest that the two Raman peaks found in laser-heated MgB$_2$ under different stress-conditions represent Raman-scattering from the same vibrations while their energies are strongly affected by deviatoric stress. The deviatoric stress in the sample aggregate is controlled by different levels of pressure gradients in the various pressure media. The strong effect of pressure gradients or rather deviatoric stress on the energy of vibrations, points to a stress-induced softmode-behavior [27]. It is important however to note here that in recent experiments performed in carbonaceous MgB$_2$, high nominal carbon concentrations induced disorder activated phonons that became stronger than the E$_{2g}$ mode at high carbon densities [28].

Fig. 2 shows the Raman spectra of the starting material and of samples retrieved from the laser-heating experiments at 9 GPa without medium and in NaCl. Absence of any shift in energy between the retrieved samples and the starting material are consistent with the non-quenchable second-order character of the proposed phase transition. Furthermore, this absence clearly argues against any change in chemistry of the laser-heated samples [29]. The energies of both peaks are the same for the starting material and the samples retrieved from high pressure and temperature, thus, the observed changes at high pressure are not induced by changes of the carbon-distribution on lattice sites or by exsolution of carbon. Both would be quenchable changes. The fact, that large incorporation of C into MgB$_2$ does not induce any structural change [30] also supports the hypothesis that the structural modification reported here is inherent to MgB$_2$.

In order to characterize further the symmetry reduction indicated by the Raman spectra of laser-heated $MgB_2$, we collected X-ray powder diffraction data. Figs. 3a and 3b show diffraction patterns of $MgB_2$ at 10 GPa after heating. The pattern in fig. 3a belongs to the sample embedded in NaCl, the other one (fig. 3b) to the sample without pressure medium. The diffraction patterns are very well matched by $MgB_2$ with space group P6/mmm. This does not come as a surprise: $MgB_2$ is a simple and compact structure. This and the bond strength of the B layers do not provide freedom for relative atomic displacements. This is indicated by the high bulk modulus [25] and by the very limited number of structural modification of the $MgB_2$ structure-type. Therefore, the observed changes in the Raman spectrum may reflect a reduction in lattice site symmetry in absence of marked atomic displacements. Within the sampled angular range we identify two additional reflections (denoted by arrows in the insets of fig. 3) of intensity at the border of the noise level at 1.47 Å and 1.25 Å which do not occur in space group P6/mmm but can be matched by $MgB_2$ cells of reduced space-group symmetries. In absence of observation of any further new Bragg peaks these two reflections are matched by a doubled $MgB_2$ cell and a reduced space-group symmetry of $P\bar{6}c2$, $P6_3cm$, $P\bar{3}c1$, or P3c1 (numbers 188, 185, 165, or 158 in the International Tables [31]). These modified $MgB_2$ cells are naturally obtained by a klassengleiche transition from P6/mmm to $P6_3mc$, along which *c* is doubled, and by subsequent reduction of symmetry by a translationsgleiche transition to cells of the mentioned space groups. Space group $P6_3mc$ itself is precluded by the observation of the reflection at 1.47 Å, which is symmetry-forbidden in this space group. In all cases the calculated structure amplitude is almost identical to that of $MgB_2$ in space group P6/mmm and the two additional peaks, being superlattice reflections, are intrinsically low in intensity. Figs. 3a and b show the refined structure models and the residuals of the refinement procedure. We used the $F_{calc}$-weighted refinement procedure in GSAS [32]. For the sample in NaCl there were 1033 observations, $\chi^2 = 22.59$, the fitted wRp was 0.0506 and Rp was 0.0242. With the background subtracted wRp was 0.1642 and Rp 0.0935. The cell dimensions of β-$MgB_2$ are *a* = 3.013(3) Å and *c* = 6.885(5) Å, which is ~ 9% shorter in *a*-direction than reported in [25] but exactly twice the dimension in *c*. This small discrepancy may result from the larger deviatoric stress and the mixed anisotropic stress-strain regime in the cold-compressed samples in [25], while the stress of the present sample-material here had been relaxed and shifted toward the Reuss-limit upon heating. The profile of the $MgB_2$ phase is matched with a small residual while the peaks of the pressure medium are much less well fitted, indicating preferred orientation after recrystallization in a temperature gradient during heating. While the pattern of $MgB_2$ heated in NaCl pressure medium exhibits rather narrow peaks and clearly shows two extra-peaks at 1.47 Å (2Θ = 14.36) and 1.25 Å (2Θ = 16.76), the pattern collected from the sample heated without pressure medium shows generally much broader peaks, indicating a lesser degree of relaxation of deviatoric stress. The refinement procedure was the same as for the sample in NaCl. The statistical parameters are $\chi^2$ = 0.646, wRp = 0.0132, Rp = 0.0070, and with background subtracted: wRp = 0.0236 and Rp = 0.0147. The fitted cell parameters are $a$ = 3.042(57) Å and $c$ = 6.797 (86) Å, which results in a 5% smaller cell than reported in [25]. There were 886 observations. The two proposed superlattice reflections at 1.25 Å (2Θ = 16.98) and 1.47 Å (2Θ = 14.59) signaling the transition to a subgroup of $P6_3mc$ are much weaker and broader in this pattern than in fig. 3a. This can be the result of the given stress-induced broadening of this peak or an intrinsic effect as the Raman spectrum of this sample also shows less pronounced splitting of the two new Raman peaks than the samples heated in pressure media (figs. 1b and c).

For convenience, we will call the proposed new phase β-$MgB_2$ and refer to magnesium diboride with space-group P6/mmm as α-$MgB_2$. In the β-phase the Mg-B and B-B distances and angles remain the same ($P\bar{6}c2$) or almost the same in $P6_3mc$, $P\bar{3}c1$, and P3c1, where the fractional $z$-coordinate of the B atoms is not fully constrained by symmetry. However, the condition of homogeneous charge distribution does not allow for itinerant changes of the $z$-coordinate (they may occur locally around defect sites or carbon-sites). In all possible space groups the site symmetry for boron is reduced with respect to the α-$MgB_2$ cell, thus constraining the vibrational degrees of freedom. This property of β-$MgB_2$ illustrates that this phase is possibly the result of a second-order transition in accordance with the softmode scenario inferred from the Raman spectra: While the entropy of the vibrational free energy is reduced, the molar volume stays almost constant during the transition. This implies two things: a) This phase is not quenchable beyond its field of thermodynamic stability. This is confirmed by the Raman spectra collected after retrieval. b) The stability field of this phase is not restricted to elevated temperature and extends to 300 K at 9 to 10 GPa. Otherwise, it would not have been observed. The occurrence after laser heating in the present experiments reflects a change in the deviatoric stress regime necessary for its formation. Stress relaxation by laser heating provides lower degrees of deviatoric stress in a material as incompressible and elastically anisotropic as $MgB_2$ than surrounding pellets of this material by soft pressure-transmitting media, which do not much affect the stress distribution inside the pellets. Further, the high-temperature treatment changes the character of the stress regime from a mixed anisotropic stress-strain regime toward the Reuss limit.

In the Raman spectrum, the differences between P6/mmm, P6$_3$mc, and its subgroups are signified by different numbers of allowed Raman active vibrations. Pure and stoichiometric $MgB_2$ in space group P6/mmm has only one Raman active vibration, which has an energy of 536 cm$^{-1}$ to 620 cm$^{-1}$ at ambient conditions [3,4,8,16] and ~ 690 to 720 cm$^{-1}$ at 9.5 GPa [17,23]. $MgB_2$ with $c$ doubled and in space group P6$_3$mc allows for two, $P\bar{3}c1$ for three, $P\bar{6}c2$ for six, P6$_3$cm for eight Raman active modes, while in P3c1 nine modes are both Raman- and IR active. Hereby we assume that the sites of the Mg and B atoms are given by the above mentioned klassen- and translationgleichen transitions, in accordance with the diffraction patterns. In general, the disappearance of the E$_{2g}$ phonon of α-$MgB_2$ and the observation of two new peaks in the Raman spectra of the transformed $MgB_2$ are consistent with the result from analysis of the diffraction patterns, since they are consistent with any of the subgroups of P6$_3$mc while this space group itself is excluded by the observation of the new diffraction peak at 1.47 Å, which is a systematically extinct reflection in this space group. As mentioned, there are distinct differences in the energies of the two Raman shifts of β-$MgB_2$ depending on the amount of deviatoric stress acting on the sample. We believe that this difference has physical relevance with respect to the electron-phonon coupling of the B-B in plane stretch vibration.

We can compare calculated energies for regular $MgB_2$, $Mg(B,C)_2$, $AlB_2$ [3] and LiBC [20,21,22] with the present measurements. In these materials the energy of B-B bond stretching vibrations varies strongly with the degree of electron-phonon coupling: E.g. in isostructural $AlB_2$ the B-B in-plane stretch vibration has an energy of 1008 cm$^{-1}$ [3], in LiBC it is at 1166 cm$^{-1}$ at ambient pressure [20, 21]. In $MgB_2$ with space group P6/mmm the extreme broadness of the Raman mode of ~ 720 cm$^{-1}$ and the fact that it has an energy about 300 to 400 cm$^{-1}$ lower than that of the corresponding vibration in isostructural and isoelectronic AlB$_2$ has been explained by strong electron phonon coupling [3]. Both spectroscopic phenomena are therefore essentially related to superconductivity in MgB$_2$ [3, 10, 15]. AlB$_2$ is isostructural and isoelectronic to regular MgB$_2$ but not superconducting [3] and does not show significant interaction of the E$_{2g}$ phonon with electrons. Consequently [3], the E$_{2g}$ phonon in AlB$_2$ is at a much higher energy (954 cm$^{-1}$ at ambient, 1045 cm$^{-1}$ at 9.5 GPa [3,15]) than in MgB$_2$ and exhibits an intense Raman peak of much smaller half-width than in MgB$_2$. This also holds true for isoelectronic LiBC, where the B-B stretching vibration of E$_g$ symmetry is shifted to 1166 cm$^{-1}$ [20, 21]. This is supported by the finding that C-containing MgB$_2$ shows hardening of the E$_{2g}$ phonon with increasing C-content while T$_c$ decreases [13, 14]. In sum the B-B stretching vibration in MgB$_2$, AlB$_2$ and LiBC ranges from 536 to 1166 cm$^{-1}$ as a function of coupling between this phonon and the B σ-electron band.

Since polarization dependent measurements are not possible for our powdered samples we cannot constrain the mode symmetry of the observed peaks. However, the proposed phase transition does not involve marked shifts of atomic positions. Hence B-B in plane stretch vibrations should not shift in energy because of changes in bond distance. In all proposed space groups, there will be Raman active B-B in plane stretching vibrations. It is therefore conceivable that the higher energetic peak in the Raman spectrum of β-MgB$_2$, which evolves from 757 to 1060 cm$^{-1}$ along with decreasing deviatoric stress is affiliated with B-B in-plane stretch and experiences different degrees of softening due to electron-phonon coupling depending on the degree of deviatoric stress acting on the sample. The other Raman-peak softens along with decreasing deviatoric stress. Hence, the Raman spectra of β-MgB$_2$ suggest that there is a mechanism of continuous structural distortion which induces decoupling of the B-B in-plane stretching vibration and the σ-electron band along with this distortion.

Before closing the discussion, one should comment on the fact that the proposed phase transition is of second order. According to this fact, the transition should also take place under compression at RT. X-ray diffraction experiments [25] on MgB$_2$ compressed at 300 K to 40 GPa did not reveal any phase transformation in MgB$_2$. Our study is different from those experiments since we used heating to high temperatures to relax stress and to shift the deviatoric stress regime toward the Reuss-limit. On the other hand, Bordet *et al.* [25] used pure $MgB_2$ as starting material, while in our study a carbonaceous material has been used. Recent Raman and $T_c$ measurements on $MgB_2$ show a kink in the pressure dependence of $T_c$ and the $E_{2g}$ phonon around 15 – 20 GPa [33, 34]. These facts were attributed to an electronic topological transition, which is consistent with the 'structurally weak', stress-triggered α-β transition proposed by us. However, it has to be confirmed that the kink in the pressure dependence of $T_c$ occurs in carbonaceous $MgB_2$ as well.

**Conclusions**

In summary, we find indications for a structural transition in carbonaceous $MgB_2$ at pressures around 9 to 10 GPa after stress relaxation by laser heating for 30 seconds to above 1500 K. Our Raman and X-ray diffraction studies suggest that this transition is of second order from P6/mmm to a subgroup of P6$_3$mc. Changes in the X-ray diffraction pattern are marginal but consistent with several subgroups of P6$_3$mc. The cell of the proposed new phase, β-$MgB_2$, is doubled with respect to the $MgB_2$ aristotype and the site symmetry of B is reduced. Consequently, the proposed new phase exhibits more than one Raman-active mode. The Raman spectrum suggests a significant reduction of electron-phonon interaction for vibrations involving B σ-bonds along with decreasing deviatoric stress, while there is a simultaneous softening of the lower energetic Raman-active vibrational mode in β-$MgB_2$ consistent with a structural softmode coupling to the lower energetic peak in the Raman spectrum of β-$MgB_2$. There is no indication for ordering of carbon on the B-sublattice. The nature of the α-β transition does not allow for quenching of the new phase from high pressures, confirmed by back-transition to the α-phase upon release. However, our work indicates that besides doping of $MgB_2$ another promising way for synthesis of similar superconductors may be application of external directed stress or internal lattice stress on materials isotopic to $MgB_2$ but of lower symmetry. A recent study on the effect of induced lattice strain in $MgB_2$ on the superconductivity [35] supports this concept. It may require extensive studies to explore the effect of deviatoric stress, non-stoichiometries, and compositional variation on the electron-phonon coupling, and to explore chemical stability of $MgB_2$ at elevated pressure. Given the experimental constraints on high-pressure experiments on materials as sensitive to deviatoric stress as $MgB_2$, the influence of small deviations from stoichiometry and of chemical impurities, a calculational approach may help to further constrain the space group of the proposed β-phase and elucidate the effect of continuous lattice distortion and domain-twinning on the electron-phonon coupling of the B-B stretch vibration. In the present study our main purpose was to provide experiment-based indications of a second order phase transition of carbonaceous $MgB_2$. Further studies to examine the effect of compositional variation and defects on the phase behaviour of this system would, all the same, be important.

**Acknowledgements:** This work was supported by the NNSA Cooperative Agreement DE-FC88-01NV14049. We acknowledge support in collecting diffraction data at sector 16 at APS by M. Somayazulu and Y. Meng. We thank G. Shen and V. Prakapenka for allowing us to use the GSECARS Raman spectrometer at APS. Use of the HPCAT and GSECARS facilities was supported by DOE-BES, DOE-NNSA, NSF, DOD -TACOM, and the W.M. Keck Foundation. Use of the APS was supported by the U.S. Department of Energy, Basic Energy Sciences, Office of Energy Research under Contract No. W-31-109-Eng-38. Daniel Errandonea acknowledges the financial support from the MCYT of Spain and the Universitat de València through the "Ramón y Cajal" program for young scientists. We greatly acknowledge the help of D.Walker in an additional experiment with a multi-anvil press.

**References**

1: J. Nagamatsu, N. Nakagawa, T. Murakana, Y. Zenitzani, and J. Akimitsu, 2001 Nature (London) **410**, 63.

2: S.L. Bud'ko *et al.*, 2001 Phys. Rev. Lett. **86**, 1877.

3: K.-P. Bohnen, R. Heid, and B. Renker, 2001 Phys. Rev. Lett. **86**, 5771.

4: J. Kortus, I.I. Mazin, K.D. Belashchenko, V.P. Antropov, and L.L. Boyer, 2001 Phys. Rev. Lett. **86**, 4656.

5: A.Y. Liu, I.I. Mazin, J. Kortus, 2001 Phys. Rev. Lett. **87**, 087005.

6: Y. Wang, T. Plackowski, and A. Junod, 2001 Physica C **355**, 179.

7: F. Bouquet, R.A. Fisher, N.E. Phillips, D.G. Hinks, and J.D. Jorgensen, 2001 Phys. Rev. Lett. **87**, 047001.

8: X.K. Chen, M.J. Konstantinovi, J.C. Irwin, D.D. Lawrie, and J.P. Franck, 2001 Phys. Rev. Lett. **87**, 157002.

9: H.J. Choi, D. Roundy, H. Sun, M.L. Cohen, and S.G. Louie. 2002 Nature **418**, 758.

10: T. Yildirim *et al.*, 2001 Phys. Rev. Lett. **87**, 037001.

11: J.D. Jorgensen, D.G.Hinks, and S. Short, 2001 Phys. Rev. B **63**, 224522.

12: S. Lee, H. Mori, T. Masui, Yu. Eltsev, A.Yamamoto, S.Tajima, 2001 J. Phys. Soc. Japan **70**, 2255.

13: T. Matsui, S. Lee, and S. Tajima, 2004 Phys. Rev. B **70**, 024504.

14: J. Arvanitidis *et al.*, 2004 J. Phys. Chem Sol. **65**, 73.

15: I. Loa, K. Kunc, K. Syassen, and P. Bouvier, 2002 Phys. Rev. B **66**, 134101.

16: J. Hlinka *et al.*, 2001, Phys. Rev. B **64**, 14503.

17: K. Kunc, I. Loa, K. Syassen, R.K. Kremer, and K. Ahn, 2001 J. Phys. Condens. Matter **13**, 9945.

18: A.P. Hammersley *et al.*, 1996 High Pres. Res. **14**, 235.

19: D. Errandonea, M. Somayazulu, D. Häusermann, and H.K. Mao, 2003 Journal of Physics: Condensed Matter **15**, 7635.

20: B. Renker *et al.*, 2003 Phys. Rev. B **69**, 057506.

21: H. Rosner, A. Kitaigorodsky, and W.E. Pickett, 2002 Phys. Rev. Lett. **88**, 127001.

22: A.M. Fogg, P.R. Chalker, J.B. Claridge, G.R. Darling, and M.J. Rosseinsky, 2003 Phys. Rev. B **67**, 245106.

23: Goncharov *et al.*, 2001 Phys. Rev. B **64**, 100509.

24: After initial heating the Raman spectrum of this sample exhibited several additional first-order Raman peaks at 242 , 345 , 447 , 553 , 698 , 912, 1013 cm$^{-1}$ besides these two peaks and the broad features around 690 and 800 cm$^{-1}$ from unconverted material while the diffraction pattern shows no difference and exhibits the same two additional Bragg-peaks at 1.47 Å and 1.25 Å also. The additional Raman-peaks disappeared upon the second heating.

25: P. Bordet *et al*., 2001 Phys. Rev. B **64**, 172502.

26: D. Errandonea, Y. Meng, M. Somayazulu, and D. Häusemann, 2005 Physica B **355**, 116.

27: G.A. Samara and P.S. Percy, 1981 Sol. State Phys. **36**, 1.

28: D. A. Tenne *et al.*, 2005 Phys. Rev. B **71**, 132512.

29: Diffraction data of material retrieved from a parallel run on $MgB_2$ + NaCl over 5 minutes in a multi-anvil press showed formation of $MgCl_2$. However, in this experiment, it was not possible to control moisture during the sample-loading process. During sample loading in DACs under a microscope attraction of water is obvious from transformation of the transparent NaCl specimen to a white, voluminous mass. We did not observe this. Further, we closed the cell after drying in a vacuum oven at 350 K for 24 hrs.

30: S. J. Balaselvi *et al.*, 2004 Physica C **407**, 31.

31: Th. Hahn, 1996 *International Tables of Crystallography*, Volume A: Space Group Symmetries, ed. Th Hahn (Kluwer International Publishers, Dordrecht, London, New York).

32: A.C. Larson and R.B. von Dreele, 1995 General Structure Analysis Software (GSAS), LAUR 86-748, Los Alamos National Laboratory, NM, USA.

33: K.P. Meletov et al., 2002 JETP Lett. **75**, 406.

34: A. F. Goncharov and V. V. Struzhkin, 2003 Physica C **385**, 117.

35: A.V. Pogrebnyakov *et al*., 2004 Phys. Rev. Lett. **93**, 147006.

**Figure Captions:**

**Figure 1a**: Raman spectra of $MgB_2$ at 10 GPa. From bottom to top we show Raman spectra of material under conditions of decreasing deviatoric stress. a) Unheated sample in a NaCl pressure medium, b) $MgB_2$ heated for 30 s to ~ 1700 K without pressure medium, c) $MgB_2$ in NaCl after heating to 1500 – 1800 K twice for 30 s, d) $MgB_2$ in Ar heated to 1700 K for 30s.

**Figure 2**: Raman spectra of $MgB_2$ at ambient conditions. The lowest spectrum was taken from the starting material, the middle spectrum is from material retrieved from the high-pressure experiment without medium, the upper one from the experiment with NaCl as pressure medium.

**Figure 3**: Powder diffraction patterns of $MgB_2$ at about 10 GPa with and without a pressure medium after laser heating. Both show the same additional peaks interpreted as superlattice reflections which characterize the proposed phase transition. Black symbols: observations, solid line: refined model, and dotted line: residual.
**a)** With NaCl as pressure medium, $\lambda = 0.3738$ Å. $MgB_2$ and NaCl reflections are indicated. The pronounced (200), (222), and (400) reflections of NaCl overlap sample peaks at 8.09, 11.45, and 16.22°. Inset: Zoom into the 2Θ range of the two proposed superlattice reflections.
**b)** Sample without pressure medium, $\lambda = 0.3681$ Å. Ticks mark the angular values of the reflections. Re gasket peaks (not fitted) are indicated. Inset: Zoom into the 2Θ range of the two proposed superlattice reflections.

**Figure 1**

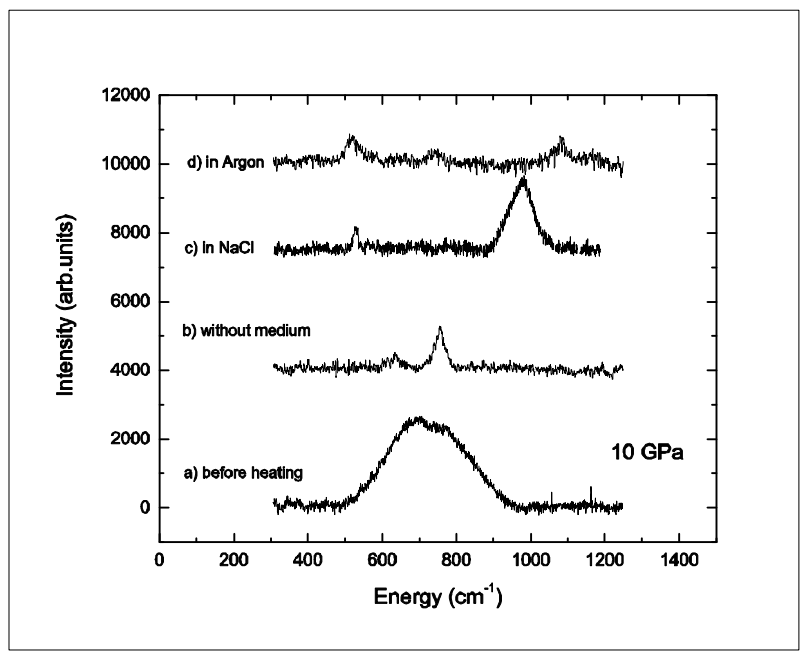

**Figure 2**

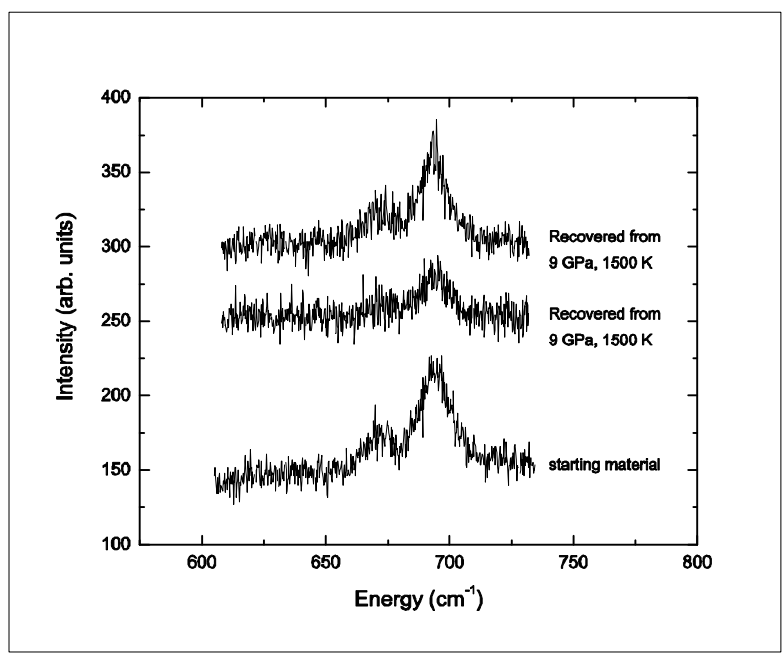
16

**Figure 3a**

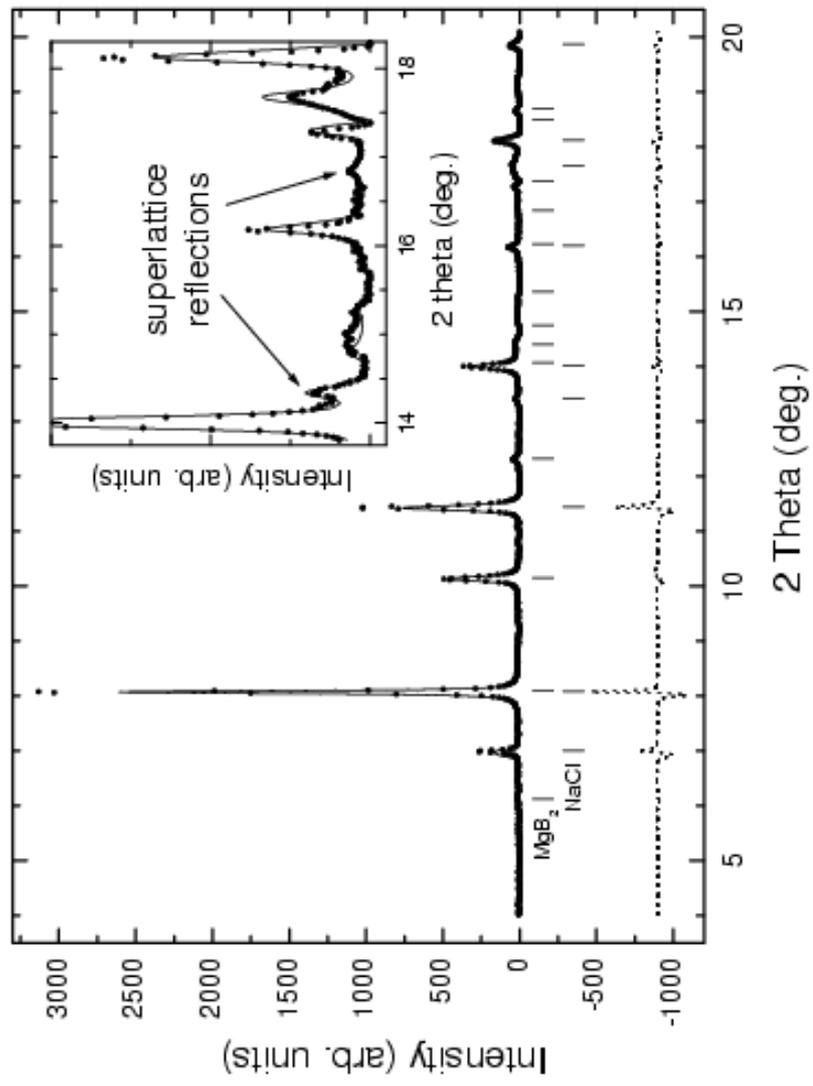

**Figure 3b**

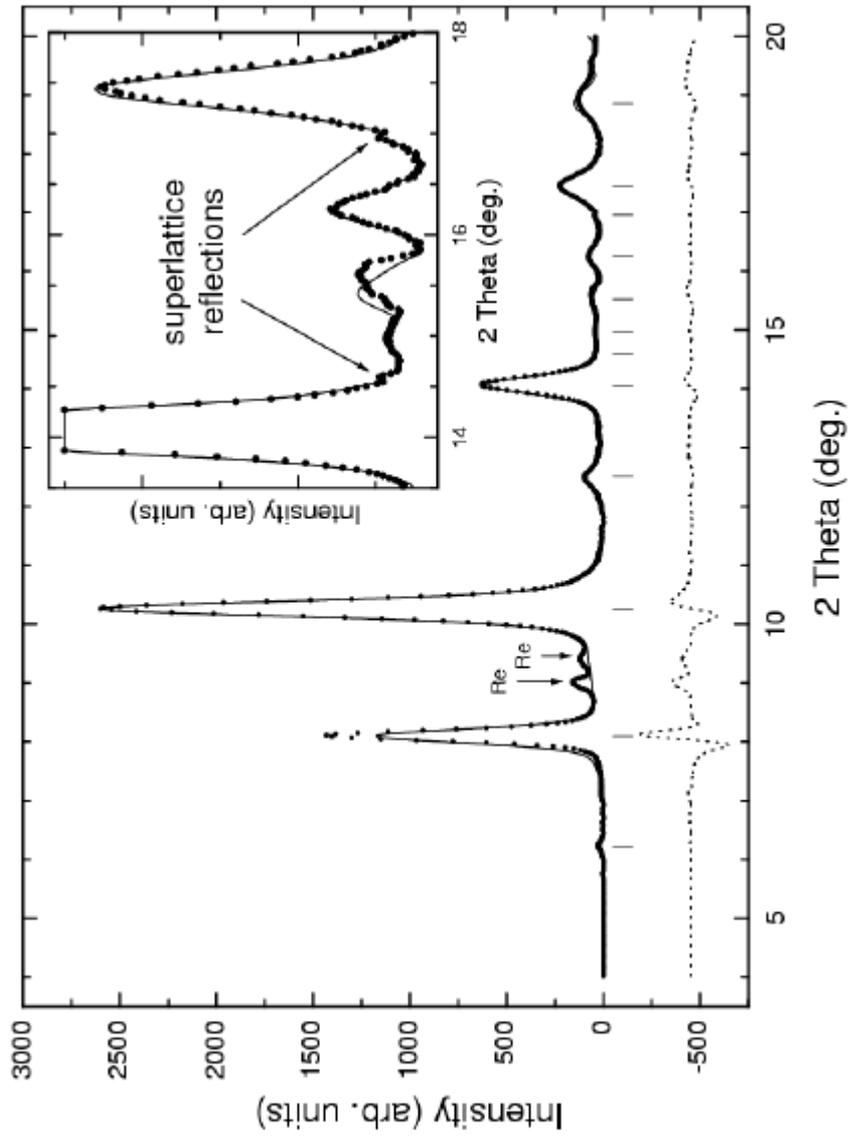